\documentclass[12pt]{article}
\usepackage{graphicx}
\usepackage[cp1251]{inputenc}

\begin{document}
 \noindent {\footnotesize\it Astronomy Letters, 2019, Vol. 45, No 9, pp. 580--592.}
 \newcommand{\dif}{\textrm{d}}

 \noindent
 \begin{tabular}{llllllllllllllllllllllllllllllllllllllllllllll}
 & & & & & & & & & & & & & & & & & & & & & & & & & & & & & & & & & & & & & \\\hline\hline
 \end{tabular}

  \vskip 0.5cm
  \centerline{\bf\large Kinematics of Hot Subdwarfs from the Gaia DR2 Catalogue}
   \bigskip
  \bigskip
  \centerline
 {V.V. Bobylev\footnote [1]{e-mail: vbobylev@gaoran.ru} and A.T. Bajkova}
  \bigskip

  \centerline{\small\it Pulkovo Astronomical Observatory, Russian Academy of Sciences,}

  \centerline{\small\it Pulkovskoe sh. 65, St. Petersburg, 196140 Russia}
 \bigskip
 \bigskip
 \bigskip

 {
{\bf Abstract}---We have studied the kinematic properties of the
candidates for hot subdwarfs (HSDs) selected by Geier et al. from
the Gaia DR2 catalogue. We have used a total of 12 515 stars with
relative trigonometric parallax errors less than 30\%. The HSDs
are shown to have different kinematics, depending on their
positions on the celestial sphere. For example, the sample of
low-latitude $(|b|<20^\circ)$ HSDs rotates around the Galactic
center with a linear velocity $V_0=221\pm5$ km s$^{-1}$. This
suggests that they belong to the Galactic thin disk. At the same
time, they lag behind the local standard of rest by $\Delta
V_\odot\sim16$ km s$^{-1}$ due to the asymmetric drift. The
high-latitude ($|b|\geq20^\circ$) HSDs rotate with a considerably
lower velocity, $V_0=168\pm6$ km s$^{-1}$. Their lagging behind
the local standard of rest is already $\Delta V_\odot\sim40$ km
s$^{-1}$. Based on the entire sample of 12 515 HSDs, we have found
a positive rotation around the $x$ axis significantly differing
from zero with an angular velocity $\omega_1=1.36\pm0.24$ km
s$^{-1}$ kpc$^{-1}$. We have studied the samples of HSDs that are
complete within $r<1.5$ kpc. Based on them, we have traced the
evolution of the parameters of the residual velocity ellipsoid as
a function of both latitude $|b|$ and coordinate $|z|.$ The
following vertical disk scale heights have been found: $h=180\pm6$
and $290\pm10$ pc from the low- and high-latitude HSDs,
respectively. A new estimate of the local stellar density
$\Sigma_{out}=53\pm4~M_{\odot}$/kpc$^2$ kpc$^{-2}$ has been
obtained for $z_{out}=0.56$ kpc from the high-latitude HSDs.
  }

 \medskip
  DOI: 10.1134/S1063773719080012

 \subsection*{INTRODUCTION}
The hot subdwarfs (sdO, sdB, sdOB) are located between the main
sequence and the white dwarf sequence on the Hertzsprung–Russell
diagram. Their peculiarity is core helium burning. Such stars were
first detected by Humason and Zwicky (1947) while studying
super-bright blue stars near the North Galactic Pole. Further
spectroscopic studies revealed a hydrogen underabundance in many
hot subdwarfs. Their temperature and surface gravity measurements
(Greenstein and Sargent 1974) allowed the proper positions of such
stars on the Hertzsprung–Russell diagram to be determined. More
specifically, they occupy a compact region at the blue end of the
horizontal branch, thereby being a late evolutionary stage of
massive stars.

The formation of hot subdwarfs (HSDs) is explained, for example,
by the loss of hydrogen by a red giant from its outer layers
before the reactions involving helium begin in the core. The cause
of this loss is not completely clear, but it is hypothesized that
this is the result of interaction between the stars in a binary
system. It is assumed that single HSDs can result from a merger of
two white dwarfs.

At present, there is extensive literature on the investigation of
single HSDs (Randall et al. 2016; Latour et al. 2018) and binary
systems with HSDs (Kupfer et al. 2015; Vos et al. 2018). Their
study can help in understanding the evolutionary peculiarities of
other galaxies, the presence of an ultraviolet excess, in
particular (Han et al. 2007), galactic globular clusters (Lei et
al. 2016), the stellar evolution (Fontaine et al. 2012), and can
help in searching for the progenitors of type Ia supernovae (Geier
et al. 2007).

Only with the appearance of highly accurate astrometric data from
the Gaia space experiment (Brown et al. 2016, 2018; Lindegren et
al. 2018) has a largescale statistical analysis of various
Galactic subsystems, in particular, a statistical and kinematic
analysis of HSDs, become possible. The methods of searching for
and identifying of such stars in largescale surveys are being
developed (Bu et al. 2017); the first big catalogues of HSDs
containing the necessary data on tens of thousands of stars have
appeared (Geier et al. 2019).

Samples of protoplanetary nebulae (PPNe) and planetary nebulae
(PNe) were studied in Bobylev and Bajkova (2017a) and Bobylev and
Bajkova (2017b), respectively. Based on three laws of the density
distribution, we determined the vertical disk scale height and
estimated the Galactic rotation parameters. Many HSDs are at a
more advanced evolutionary stage compared to PPNe and PNe, which
is important for studying the various stages of stellar evolution.

The goal of this paper is to analyze the kinematics and spatial
distribution of a large sample of HSD candidates. Such a sample of
39 800 stars has recently been produced by Geier et al. (2019)
using data from the Gaia DR2 catalogue (Brown et al. 2018;
Lindegren et al. 2018) and a number of photometric sky surveys.

 \section*{DATA}
In this paper we use the catalogue by Geier et al. (2019). It
contains 39 800 HSD candidates selected from the Gaia DR2
catalogue in combination with several ground-based multiband
photometric sky surveys. HSDs of spectral types O and B, late- B
stars in the blue part of the horizontal branch, hot post-AGB
stars, and PN central stars are expected to constitute the
majority of candidates. The authors of the catalogue believe that
the contamination by cooler stars is about 10\%. The selected HSDs
are distributed over the entire celestial sphere. According to the
estimates by Geier et al. (2019), the catalogue is complete to a
distance of 1.5 kpc, except for the narrow region near the
Galactic plane and the zones with the Magellanic Clouds.

For each star this catalogue gives the trigonometric parallax
$\pi$ and two proper motion components, $\mu_\alpha\cos \delta$
and $\mu_\delta$. There are no radial velocities. Extensive
photometric information is also available.

Of course, the radial velocities have been measured for several
hundred HSDs. Note, for example, the catalogue of radial
velocities for such stars in close binary systems by Geier et al.
(2015) or Kupfer et al. (2015). In those cases where it is
possible to construct the orbit of a binary or multiple system,
the systemic velocity $V_\gamma$ is usually determined with an
error of 1--5 km s$^{-1}$. However, for such a large number as 39
800 HSDs no highly accurate radial velocities have been measured
so far.

In this paper, to determine the kinematic parameters, we took
stars with relative trigonometric parallax errors less than 30\%.
In addition, the zero-point correction $\Delta\pi$ was added to
the Gaia DR2 parallaxes; thus, the new stellar parallax is
$\pi+0.050$ mas. The necessity of applying this systematic
correction $\Delta\pi=-0.029$ mas (here the minus means that the
correction should be added to the Gaia DR2 stellar parallaxes to
reduce them to the standard) was first pointed out by Lindegren et
al. (2018) and confirmed by Arenou et al. (2018). Slightly later,
having analyzed various highly accurate sources of distance
scales, Stassun and Torres (2018), Riess et al. (2018), Zinn et
al. (2018), and Yalyalieva et al. (2018) showed the correction to
be $\Delta\pi=-0.050$ mas with an error of several units of the
last decimal digit.

 \section*{METHODS}
In this paper we consider stars only with the proper motions,
because there is no information about the radial velocities in the
catalogue by Geier et al. (2019). Thus, two tangential velocity
components are known from observations, $V_l=4.74r\mu_l\cos b$ and
$V_b=4.74r\mu_b$ along the Galactic longitude $l$ and latitude
$b,$ respectively, expressed in km s$^{-1}$. Here, the coefficient
4.74 is the ratio of the number of kilometers in an astronomical
unit to the number of seconds in a tropical year, and $r=1/\pi$ is
the stellar heliocentric distance in kpc that we calculate via the
stellar parallax $\pi.$ The proper motion components $\mu_l\cos b$
and $\mu_b$ are expressed in mas yr$^{-1}$.

To determine the parameters of the Galactic rotation curve, we use
the equations derived from Bottlinger's formulas in which the
angular velocity $\Omega$ is expanded into a series to terms of
the second order of smallness in $r/R_0:$
 \begin{equation}
 \begin{array}{lll}
 V_l= U_\odot\sin l-V_\odot\cos l +r\Omega_0\cos b\\
 -(R-R_0)(R_0\cos l-r\cos b)\Omega^\prime_0
 -0.5(R-R_0)^2(R_0\cos l-r\cos b)\Omega^{\prime\prime}_0\\
 -r\cos l\sin b\omega_1
 -r\sin l\sin b\omega_2,
 \label{EQ-2}
 \end{array}
 \end{equation}
 \begin{equation}
 \begin{array}{lll}
 V_b=U_\odot\cos l\sin b + V_\odot\sin l \sin b -W_\odot\cos b\\
    +R_0(R-R_0)\sin l\sin b\Omega^\prime_0
 +0.5R_0(R-R_0)^2\sin l\sin b\Omega^{\prime\prime}_0\\
 +r\sin l\omega_1-r\cos l\omega_2,
 \label{EQ-3}
 \end{array}
 \end{equation}
where $R$ is the distance from the star to the Galactic rotation
axis (cylindrical radius vector):
  \begin{equation}
 R^2=r^2\cos^2 b-2R_0 r\cos b\cos l+R^2_0.
 \end{equation}
$\Omega_0$ is the angular velocity of Galactic rotation at the
solar distance $R_0,$ the parameters $\Omega^{\prime}_0$ and
$\Omega^{\prime\prime}_0$ are the corresponding derivatives of the
angular velocity, and $V_0=|R_0\Omega_0|$. Apart from the rotation
around the Galactic $z$ axis (described by the parameter
$\Omega$), in this paper we also consider the angular velocities
of rotation around the $x$ and $y$ axes described by the
parameters $\omega_1$ and $\omega_2$, respectively.

Note that in this paper the signs of the unknowns are taken in
such a way that the positive rotations occur from the $x$ axis to
$y,$ from $y$ to $z,$ and from $z$ to $x.$ Thus, the angular
velocity of Galactic rotation will be negative, in contrast to
many other papers (Rastorguev et al. 2017; Vityazev et al. 2017;
Bobylev and Bajkova 2017a), where, for convenience, it is deemed
positive.

The interest in $\omega_1$ and $\omega_2$ stems from the following
factors. First, this can be a slight residual rotation of the Gaia
DR2 reference frame relative to the inertial reference frame.
Second, this can be a reflection of the Galactic warp or other
processes causing the vertical oscillations of stars in the
Galaxy. A uniform distribution of the stars being analyzed over
the celestial sphere is required for a reliable determination of
the parameters $\omega_1$ and $\omega_2$. Dwarfs and subdwarfs are
well suited for solving this problem.

In the system of conditional equations (1)--(2) eight unknowns are
to be determined: $U_\odot,$ $V_\odot,$ $W_\odot,$
 $\Omega_0,$ $\Omega^\prime_0,$ $\Omega^{\prime\prime}_0,$
 $\omega_1$ and $\omega_2$. We find
their values by solving the conditional equations by the
least-squares method (LSM). Weights of the form $w_l=S_0/\sqrt
{S_0^2+\sigma^2_{V_l}}$ and $w_b=S_0/\sqrt
{S_0^2+\sigma^2_{V_b}},$ are used, where $S_0$ is the ``cosmic''
dispersion (for each sample we find its value in advance to be
close to the error per unit weight $\sigma_0,$ obtained by
presolving the equations), $\sigma_{V_l}$ and $\sigma_{V_b}$ are
the errors in the corresponding observed velocities. $S_0$ is
comparable to the root-mean-square residual $\sigma_0$ (the error
per unit weight) calculated by solving the conditions equations
(1)--(2). The solution is sought in several iterations by applying
the $3\sigma$ criterion to exclude the stars with large residuals.

Note several papers devoted to determining the mean Galactocentric
distance of the Sun using its individual determinations made in
the last decade by independent methods. For example,
$R_0=8.0\pm0.2$ kpc (Valle\'e 2017), $R_0=8.40\pm0.4$ kpc (de
Grijs and Bono 2017) or $R_0=8.0\pm0.15$ kpc (Camarillo et al.
2018). Based on these reviews, in this paper we adopt
$R_0=8.0\pm0.15$ kpc.

 \subsection*{Residual Velocity Ellipsoid}
We estimated the dispersion of the stellar residual velocities
using the following well-known method (Ogorodnikov 1965). Let
$U,V,W$ be the velocities along the coordinate axes $x, y,$ and
$z.$ Let us consider the six second-order moments $a, b, c, f, e,
d:$
\begin{equation}
 \begin{array}{lll}
 a=\langle U^2\rangle-\langle U^2_\odot\rangle,\\
 b=\langle V^2\rangle-\langle V^2_\odot\rangle,\\
 c=\langle W^2\rangle-\langle W^2_\odot\rangle,\\
 f=\langle VW\rangle-\langle V_\odot W_\odot\rangle,\\
 e=\langle WU\rangle-\langle W_\odot U_\odot\rangle,\\
 d=\langle UV\rangle-\langle U_\odot V_\odot\rangle,
 \label{moments}
 \end{array}
 \end{equation}
which are the coefficients of the equation for the surface
 \begin{equation}
 ax^2+by^2+cz^2+2fyz+2ezx+2dxy=1,
 \end{equation}
and also the components of the symmetric tensor of moments of the
residual velocities:
 \begin{equation}
 \left(\matrix {
  a& d & e\cr
  d& b & f\cr
  e& f & c\cr }\right).
 \label{ff-5}
 \end{equation}
To determine the values in this tensor in the absence of
radial-velocity data, the following three equations are used:
\begin{equation}
 \begin{array}{lll}
 V^2_l= a\sin^2 l+b\cos^2 l\sin^2 l-2d\sin l\cos l,
 \label{EQsigm-2}
 \end{array}
 \end{equation}
\begin{equation}
 \begin{array}{lll}
 V^2_b= a\sin^2 b\cos^2 l+b\sin^2 b\sin^2 l+c\cos^2 b\\
 -2f\cos b\sin b\sin l-2e\cos b\sin b\cos l+2d\sin l\cos l\sin^2 b,
 \label{EQsigm-3}
 \end{array}
 \end{equation}
\begin{equation}
 \begin{array}{lll}
 V_lV_b= a\sin l\cos l\sin b+b\sin l\cos l\sin b\\
 +f\cos l\cos b-e\sin l\cos b+d(\sin^2 l\sin b-\cos^2\sin b),
 \label{EQsigm-4}
 \end{array}
 \end{equation}
which are solved by the least-squares method for the six unknowns
$a,b,c,f,e,$ and $d$. The eigenvalues of the tensor (6)
$\lambda_{1,2,3}$ are then found from the solution of the secular
equation
 \begin{equation}
 \left|\matrix
 {
a-\lambda&          d&        e\cr
       d & b-\lambda &        f\cr
       e &          f&c-\lambda\cr
 }
 \right|=0.
 \label{ff-7}
 \end{equation}
The eigenvalues of this equation are equal to the reciprocals of
the squares of the semiaxes of the velocity moment ellipsoid and,
at the same time, the squares of the semiaxes of the residual
velocity ellipsoid:
 \begin{equation}
 \begin{array}{lll}
 \lambda_1=\sigma^2_1, \lambda_2=\sigma^2_2, \lambda_3=\sigma^2_3,\qquad
 \lambda_1>\lambda_2>\lambda_3.
 \end{array}
 \end{equation}
We found the directions of the main axes of the tensor (10),
$L_{1,2,3}$ and $B_{1,2,3},$ from the relations:
 \begin{equation}
 \tan L_{1,2,3}={{ef-(c-\lambda)d}\over {(b-\lambda)(c-\lambda)-f^2}},
 \label{ff-41}
 \end{equation}
 \begin{equation}
 \tan B_{1,2,3}={{(b-\lambda)e-df}\over{f^2-(b-\lambda)(c-\lambda)}}\cos L_{1,2,3}.
 \label{ff-42}
 \end{equation}

 \subsection*{Exponential Density Distribution}
In the case of an exponential distribution of the density $\rho$
along the $z$ coordinate axis, we have
 \begin{equation}
  \rho=\rho_0 \exp \biggl(-{|z-z_\odot|\over h} \biggr),
 \label{rho}
 \end{equation}
where $\rho_0$ is the normalization constant; $z_\odot$ is the
mean value calculated from the $z$ coordinates of the sample
stars, which reflects the well-known fact of the Sun's elevation
above the Galactic plane; and $h$ is the vertical scale height.
The observed histogram of the distribution of stars along the $z$
axis is described by an analogous expression:
 \begin{equation}
  N(z)=N_0 \exp \biggl(-{|z-z_\odot|\over h} \biggr),
 \label{exp}
 \end{equation}
where $N_0$ is the normalization coefficient.

The surface density of gravitating matter $\Sigma_{out}(z_{out})$
within a distance $z_{out}$ from the Galactic plane $z=0$ can be
found from the following relation:
\begin{equation}
 \begin{array}{lll}
 \renewcommand{\arraystretch}{3.2}
 \displaystyle
       \qquad\qquad\qquad\Sigma_{out}(z_{out})=\\
 \displaystyle
 -\frac{\overline {v^2_z}}{2\pi G}
  \Bigg{(} {\frac{1}{\rho}}{\frac{\partial\rho}{\partial z}} \Bigg{)}\Bigg{|}_{z_{out}}
 +\frac{2z_{out}(B^2-A^2)}{2\pi G},
\label{Sigma}
 \end{array}
\end{equation}
where the gravitational constant $G$ is taken to be equal to one.
Following the approach by Korchagin et al. (2003) based on the
solution by Spitzer (1942) for a self-gravitating disk, we have
 $$
 \Bigg{(} {\frac{1}{\rho}}{\frac{\partial\rho}{\partial z}}
 \Bigg{)}\Bigg{|}_{z_{out}}=h.
 $$
Given $z_{out}$, the mean vertical velocity $\overline {v^2_z},$
the scale height $h,$ and the Oort constants $A$ and $B,$ we can
estimate the surface density from the relation
\begin{equation}
 \renewcommand{\arraystretch}{2.2}
 \Sigma_{out}(z_{out})=
 -\frac{\overline {v^2_z} \cdot h}{2\pi G}+\frac{2z_{out}(B^2-A^2)}{2\pi G}.
\end{equation}

 \begin{table}[p]
 \caption[]{\small
Galactic rotation parameters found from the stars with relative
trigonometric parallax errors less than 15\% in the upper part of
the table and less than 30\% in its lower part
 }
  \begin{center}  \label{t:01}
  \small
  \begin{tabular}{|l|r|r|r|r|r|}\hline
    Parameters                   &   All stars    & $|b|<20^\circ$  & $|b|\geq20^\circ$  \\\hline
    $U_\odot,$    km s$^{-1}$    &  $11.71\pm0.59$ &  $10.31\pm0.65$ &  $12.80\pm1.05$ \\
    $V_\odot,$    km s$^{-1}$    &  $38.51\pm0.90$ &  $28.79\pm1.16$ &  $47.36\pm1.43$ \\
    $W_\odot,$    km s$^{-1}$    &  $ 6.17\pm0.54$ &  $ 7.19\pm0.52$ &  $ 4.32\pm1.15$ \\
   $\Omega_0,$    km s$^{-1}$ kpc$^{-1}$ & $-26.28\pm0.52$ & $-28.68\pm0.52$ & $-23.65\pm1.03$ \\
  $\Omega^{'}_0,$ km s$^{-1}$ kpc$^{-2}$ & $  2.68\pm0.13$ & $  3.46\pm0.14$ & $  2.35\pm0.23$ \\
 $\Omega^{''}_0,$ km s$^{-1}$ kpc$^{-3}$ & $  0.02\pm0.17$ & $ -0.60\pm0.21$ & $  0.97\pm0.29$ \\
      $\omega_1,$ km s$^{-1}$ kpc$^{-1}$ & $ 1.20\pm0.38$  & $ 0.91\pm0.43$  & $ 1.58\pm0.65$ \\
      $\omega_2,$ km s$^{-1}$ kpc$^{-1}$ & $-0.06\pm0.45$  & $ 0.41\pm0.57$  & $ 0.45\pm0.74$ \\
      $\sigma_0,$ km s$^{-1}$    &            39.6 &            30.5 &           50.2 \\
      $N_\star$                  &            7128 &            3767 &           3361 \\
      ${\overline r},$ kpc       &            1.41 &            1.35 &           1.48 \\
              $A$ km s$^{-1}$ kpc$^{-1}$ & $ 10.75\pm0.51$ & $ 13.86\pm0.57$ & $  9.41\pm0.92$ \\
              $B$ km s$^{-1}$ kpc$^{-1}$ & $-15.53\pm0.72$ & $-14.82\pm0.78$ & $-14.24\pm1.38$ \\
            $V_0$ km s$^{-1}$    & $   210\pm6$    & $   229\pm6$    & $   189\pm9$    \\
  \hline
    $U_\odot,$    km s$^{-1}$    &  $12.33\pm0.51$  &  $11.61\pm0.54$  &  $12.11\pm0.90$ \\
    $V_\odot,$    km s$^{-1}$    &  $43.14\pm0.72$  &  $28.10\pm0.84$  &  $54.80\pm1.19$ \\
    $W_\odot,$    km s$^{-1}$    &  $ 6.15\pm0.47$  &  $ 7.07\pm0.45$  &  $ 4.02\pm0.98$ \\
   $\Omega_0,$    km s$^{-1}$ kpc$^{-1}$ & $-24.45\pm0.32$  & $-27.66\pm0.31$  & $-20.96\pm0.65$ \\
  $\Omega^{'}_0,$ km s$^{-1}$ kpc$^{-2}$ & $  2.57\pm0.08$  & $  3.27\pm0.08$  & $  1.91\pm0.15$ \\
 $\Omega^{''}_0,$ km s$^{-1}$ kpc$^{-3}$ & $-0.127\pm0.066$ & $-0.427\pm0.069$ & $ 0.692\pm0.121$ \\
      $\omega_1,$ km s$^{-1}$ kpc$^{-1}$ & $ 1.36\pm0.24$   & $ 0.51\pm0.26$   & $ 1.90\pm0.41$ \\
      $\omega_2,$ km s$^{-1}$ kpc$^{-1}$ & $-0.53\pm0.27$   & $-0.03\pm0.33$   & $-0.22\pm0.45$ \\
      $\sigma_0,$ km s$^{-1}$    &            44.1 &            33.1 &           55.9 \\
      $N_\star$                  &           12515 &            6690 &           5825 \\
      ${\overline r},$ kpc       &            1.90 &            1.83 &           1.97 \\
              $A$ km s$^{-1}$ kpc$^{-1}$ & $ 10.26\pm0.31$ & $ 13.08\pm0.32$ & $  7.64\pm0.59$ \\
              $B$ km s$^{-1}$ kpc$^{-1}$ & $-14.18\pm0.45$ & $-14.58\pm0.45$ & $-13.32\pm0.88$ \\
            $V_0$ km s$^{-1}$    & $   196\pm5$    & $   221\pm5$    & $   168\pm6$    \\
  \hline
 \end{tabular}\end{center}
  {\small $N_\star$ is the number of stars used,
  ${\overline r}$ is the mean distance of the sample of stars,
  and $\sigma_0$ is the error per unit weight.}\end{table}

 \section*{RESULTS AND DISCUSSION}
 \subsection*{Galactic Rotation Parameters}
We have run into the fact that the HSDs under consideration have
different kinematic properties, depending on their positions in
the Galaxy. Therefore, the entire sample was divided into two
groups, depending on the absolute value of the Galactic latitude:
low-latitude $(|b|<20^\circ)$ and high-latitude
$(|b|\geq20^\circ)$ stars.

The kinematic parameters found for these stars are given in Table
1. Apart from the eight sought for kinematic parameters, it gives
the following: the mean distance of the sample of stars
${\overline r}$, the error per unit weight $\sigma_0$ that we
estimate when seeking the LSM solution of the system of
conditional equations (1)--(2), the Oort constants $A$ and $B$
that we calculate based on the following relations:
\begin{equation}
 A= 0.5\Omega^{'}_0R_0,\quad
 B=-\Omega_0+A,
 \label{AB}
\end{equation}
and the linear rotation velocity of the Galaxy at the solar
distance $V_0=|R_0\Omega_0|$. When calculating the kinematic
parameters presented in Table 1, we took the stars from the
interval 0.5--6 kpc and rejected the stars with large (more than
60 km s$^{-1}$) random measurement errors of their proper motions.

The separation in relative parallax error was made for the
following reasons. On the one hand, using the stars with relative
trigonometric parallax errors less than 30\% allows us to use a
large number of objects and to estimate the sought-for parameters
with smaller errors. On the other hand, the sample of stars with
parallax errors less than 15\% allows more local parameters, in
particular, more reliable estimates of the velocities
$(U,V,W)_\odot$ and $\Omega_0$ to be obtained.

First note the parameters that describe the kinematics of the most
rapidly rotating Galactic subsystems. For example, based on 130
masers with measured VLBI trigonometric parallaxes, Rastorguev et
al. (2017) found the following quantities:
 $(U,V)_\odot=(11.40,17.23)\pm(1.33,1.09)$ km s$^{-1}$,
 $\Omega_0=-28.93\pm0.53$ km s$^{-1}$ kpc$^{-1}$,
 $\Omega^{'}_0=3.96\pm0.07$ km s$^{-1}$ kpc$^{-2}$,
$\Omega^{''}_0=-0.87\pm0.03$ km s$^{-1}$ kpc$^{-3}$, and
 $V_0=243\pm10$ km s$^{-1}$ (for the derived $R_0=8.40\pm0.12$ kpc).
Based on a sample of young $(\log t<8)$ open star clusters with
the proper motions and distances from the Gaia DR2 catalogue,
Bobylev and Bajkova (2019) obtained the following estimates:
 $(U,V,W)_\odot=(8.53,11.22,7.83)\pm(0.38,0.46,0.32)$ km s$^{-1}$,
  $\Omega_0=-28.71\pm0.22$ km s$^{-1}$ kpc$^{-1}$,
  $\Omega^{'}_0= 4.100\pm0.058$ km s$^{-1}$ kpc$^{-2}$, and
  $\Omega^{''}_0=-0.736\pm0.033$ km s$^{-1}$ kpc$^{-3}$, where
  $V_0=230\pm5$ km s$^{-1}$ (for the adopted $R_0=8.0\pm0.15$ kpc).

The linear velocity $V_0=229\pm6$ km s$^{-1}$ inferred from the
low-latitude HSDs (the upper part of Table 1) suggests that this
population of stars belongs to the thin disk. At the same time,
the velocity $V_\odot\sim28$ km s$^{-1}$ derived from them shows
that they lag behind the local standard of rest by $\sim16$ km
s$^{-1}$ due to the so-called asymmetric drift. One of the
reliable present-day determinations of the parameters of the
peculiar solar motion relative to the local standard of rest was
made by Sch\"onrich et al. (2010),
$(U_\odot,V_\odot,W_\odot)=(11.1,12.2,7.3)\pm(0.7,0.5,0.4)$ km
s$^{-1}$. For all of the older Galactic objects their lagging
behind the local standard of rest increases, the velocity
$V_\odot$ rises, due to the asymmetric drift. As can be seen from
the last column in the lower part of Table 1, the lagging of the
high-latitude HSDs behind the local standard of rest is already
more than 40 km s$^{-1}$.

The first and second derivatives of the angular velocity
$\Omega^{'}_0$ and $\Omega^{''}_0$ are also determined
satisfactorily (in agreement with our analysis of young thin-disk
objects) from the low-latitude HSDs. Nevertheless, the absolute
value of the Oort constant $A$ is less than that of the Oort
constant B for these stars as well, suggesting that the linear
rotation velocity in the neighborhood increases with $R.$ For the
youngest Galactic objects, for example, maser sources or OB stars,
the reverse situation is a typical one, i.e., the absolute value
of the Oort constant $A$ is greater than that of the Oort constant
$B$ (Vityazev et al. 2017; Bobylev and Bajkova 2014, 2019).

The second derivative $\Omega^{''}_0$ is not determined from the
high-latitude HSDs, while the first derivative $\Omega^{'}_0$
differs greatly from its value found from younger objects.

 \subsection*{Rotation around the $x$ and $y$ Axes}
From the data in Table 1 we can unambiguously conclude that there
is no rotation around the $y$ axis ($\omega_2$) differing
significantly from zero in our samples.

In contrast, there is a rotation around the $x$ axis ($\omega_1$)
differing significantly from zero in the samples of high-latitude
HSDs and the ``all stars'' sample. The nature of this rotation has
not yet been established. First, this can be interpreted as a
residual rotation of the Gaia DR2 reference frame relative to the
reference frame of extragalactic sources. In this case,
$\omega_1=1.36\pm0.24$ km s$^{-1}$ kpc$^{-1}$ from the first
column in the lower part of Table 1 found from all stars is of
value. Taking into account the mean distance of this sample, we
obtain $\omega_1=0.15\pm0.03$ mas yr$^{-1}$. Second, this can be a
manifestation of some large-scale physical process characteristic
of the high-latitude HSDs.

Note also that the angular velocities $\omega_1$ and $\omega_2$
cannot be determined reliably from the low-latitude objects,
because $\sin b$ in Eq. (1) is close to zero. Therefore, the
parameters $\omega_1$ and $\omega_2$ are actually found only from
one Eq. (2).

Lindegren et al. (2018) concluded that the Gaia DR2 reference
frame has no rotation relative to the reference frame of quasars
within 0.15 mas yr$^{-1}$ (along three axes), with the effect
being most pronounced in the region of bright $(G<12^m)$ stars.

Having analyzed $\sim$6 million stars from the Gaia DR2 catalogue,
Tsvetkov and Amosov (2019) found $\omega_1\sim0.7\pm0.1$1 km
s$^{-1}$ kpc$^{-1}$ significantly differing from zero, which is
stably preserved depending on the mean distance of the sample.
When studying $\sim$900 open star clusters with data from the Gaia
DR2 catalogue, Bobylev and Bajkova (2019) detected a rotation of
this entire sample around the Galactic $x$ axis with an angular
velocity $\omega_1=0.48\pm0.15$ km s$^{-1}$ kpc$^{-1}$.

The idea that the kinematics of the observed stars is affected by
the large-scale Galactic warp underlies the second approach. Based
on the simplest solid-body rotation model, for example, Miyamoto
and Zhu (1998) found a rotation of this system of stars around the
Galactic x axis with an angular velocity of about 4 km s$^{-1}$
kpc$^{-1}$ from the proper motions of O--B5 stars. Bobylev (2010)
found a rotation of this system of stars around the $x$ axis with
an angular velocity of about $-4$ km s$^{-1}$ kpc$^{-1}$ from the
proper motions of $\sim$80 000 red giant clump stars. Bobylev
(2013) found a rotation around the $x$ axis with an angular
velocity of $-15\pm5$ km s$^{-1}$ kpc$^{-1}$ from Cepheids. On the
whole, the observations confirm the asymmetry in the vertical
stellar velocities (L\'opez-Corredoira et al. 2014; Romero-G\'omez
et al. 2018), but the application of a more complex disk
precession model is required to describe the phenomenon.

 \subsection*{Parameters of the Residual Velocity Ellipsoid}
To form the moments (4) for each subsample, the corresponding
velocities $U_\odot, V_\odot,$ and $W_\odot$ were taken from Table
1. In addition, the stellar velocities were corrected for the
Galactic rotation. For this purpose, we use the parameters of the
rotation curve derived from the low-latitude HSDs (the upper part
of Table 1).

 \begin{table}[t]
 \caption[]
  {\small Parameter $z_\odot$ and $h$}
  \begin{center}  \label{t:02}
  \begin{tabular}{|l|c|c|c|c|c|}\hline
 Objects          & $N_\star$ & $z_\odot,$~pc & $h,$~pc    & Reference           \\\hline

 HSDs,    $|b|<20^\circ$  &  6690 &  $-45\pm5~$ & $190\pm4~$ & This paper \\
 HSDs, $|b|\geq20^\circ$  &  5825 &  $-~5\pm17$ & $700\pm8~$ & This paper \\\hline
 Cepheids, $t\sim75~$ Myr &   246 &  $-23\pm2~$ & $~70\pm2~$ & (5) \\
 Cepheids, $t\sim138$ Myr &   250 &  $-24\pm2~$ & $~84\pm2~$ & (5) \\
 PPNe, sample 1           &   107 &  $-28\pm12$ & $146\pm15$ & (1) \\
 OSCs, 200--1000 Myr      &   148 &  $-15\pm2~$ & $150\pm27$ & (3) \\
 PN                       &   230 &  $-~6\pm7~$ & $197\pm10$ & (2) \\
 White dwarfs             &   717 &        0~~  &  220--300  & (4) \\
 PPNe, sample 2           &   ~88 &  $-37\pm53$ & $568\pm42$ & (1) \\
 Disk HSDs                &   114 &        0~~  & $930\pm90$ & (6) \\
 \hline
 HSDs,    $|b|<20^\circ,$ $r<1.5$~kpc &  3118 &  $-15\pm4~$ & $180\pm6~$ & This paper \\
 HSDs, $|b|\geq20^\circ,$ $r<1.5$~kpc &  2363 &  $-46\pm13$ & $290\pm10$ & This paper \\
 \hline
  \end{tabular}\end{center}  {\small
$N_\star$ is the number of objects, PPNe stands for protoplanetary
nebulae, PNe stands for planetary nebulae, HSDs stands for HSDs,
(1)—Bobylev and Bajkova (2017a), (2)—Bobylev and Bajkova (2017b),
(3)—Bonatto et al. (2006), (4)—Vennes et al. (2002), (5)—Bobylev
and Bajkova (2016), (6)—Altmann et al. (2004).
 }
 \end{table}

Based on our sample of 4181 low-latitude stars with relative
trigonometric parallax errors less than 15\%, we found the
following residual velocity dispersions:
 \begin{equation}
 \begin{array}{lll}
  \sigma_1=37.4\pm0.9~\hbox{km s$^{-1}$}, \\
  \sigma_2=28.1\pm0.7~\hbox{km s$^{-1}$}, \\
  \sigma_3=22.8\pm0.9~\hbox{km s$^{-1}$},
 \label{rezult-3}
 \end{array}
 \end{equation}
and the following orientation parameters of this ellipsoid:
 \begin{equation}
  \matrix {
  L_1=~~3\pm4^\circ, & B_1=~1\pm4^\circ, \cr
  L_2=~93\pm3^\circ, & B_2=-2\pm3^\circ, \cr
  L_3=127\pm3^\circ, & B_3=88\pm4^\circ. \cr
   }
 \label{rezult-33}
 \end{equation}
Obviously, the orientation of this residual velocity ellipsoid
coincides with the $x, y, z$ coordinate axes with a high accuracy.
From the low-latitude HSDs, but with relative trigonometric
parallax errors less than 30\%(7156 stars) we found the following
residual velocity dispersions:
 \begin{equation}
 \begin{array}{lll}
  \sigma_1=38.0\pm0.6~\hbox{km s$^{-1}$}, \\
  \sigma_2=32.6\pm0.8~\hbox{km s$^{-1}$}, \\
  \sigma_3=25.7\pm0.6~\hbox{km s$^{-1}$},
 \label{rezult-4}
 \end{array}
 \end{equation}
while the orientation of this ellipsoid is
 \begin{equation}
  \matrix {
  L_1=~15\pm8^\circ, & B_1=~3\pm1^\circ, \cr
  L_2=105\pm4^\circ, & B_2=-2\pm2^\circ, \cr
  L_3=156\pm4^\circ, & B_3=87\pm3^\circ. \cr
   }
 \label{rezult-44}
 \end{equation}
The orientation of this ellipsoid is also close to the directions
of the $x, y$ and $z$ coordinate axes; only $L_1$ is determined
with a large error.

Based on our sample of 3584 high-latitude HSDs with relative
trigonometric parallax errors less than 15\%, we found the
following residual velocity dispersions:
 \begin{equation}
 \begin{array}{lll}
  \sigma_1=51.9\pm1.1~\hbox{km s$^{-1}$}, \\
  \sigma_2=46.6\pm1.8~\hbox{km s$^{-1}$}, \\
  \sigma_3=34.8\pm0.8~\hbox{km s$^{-1}$},
 \label{rezult-5}
 \end{array}
 \end{equation}
and the orientation parameters of this ellipsoid
 \begin{equation}
  \matrix {
  L_1=~13\pm6^\circ~, & B_1=~-1\pm1^\circ, \cr
  L_2=103\pm10^\circ, & B_2=-15\pm2^\circ, \cr
  L_3=~98\pm3^\circ~, & B_3=~75\pm4^\circ. \cr
   }
 \label{rezult-55}
 \end{equation}

 \begin{figure} {\begin{center}
 \includegraphics[width=140mm]{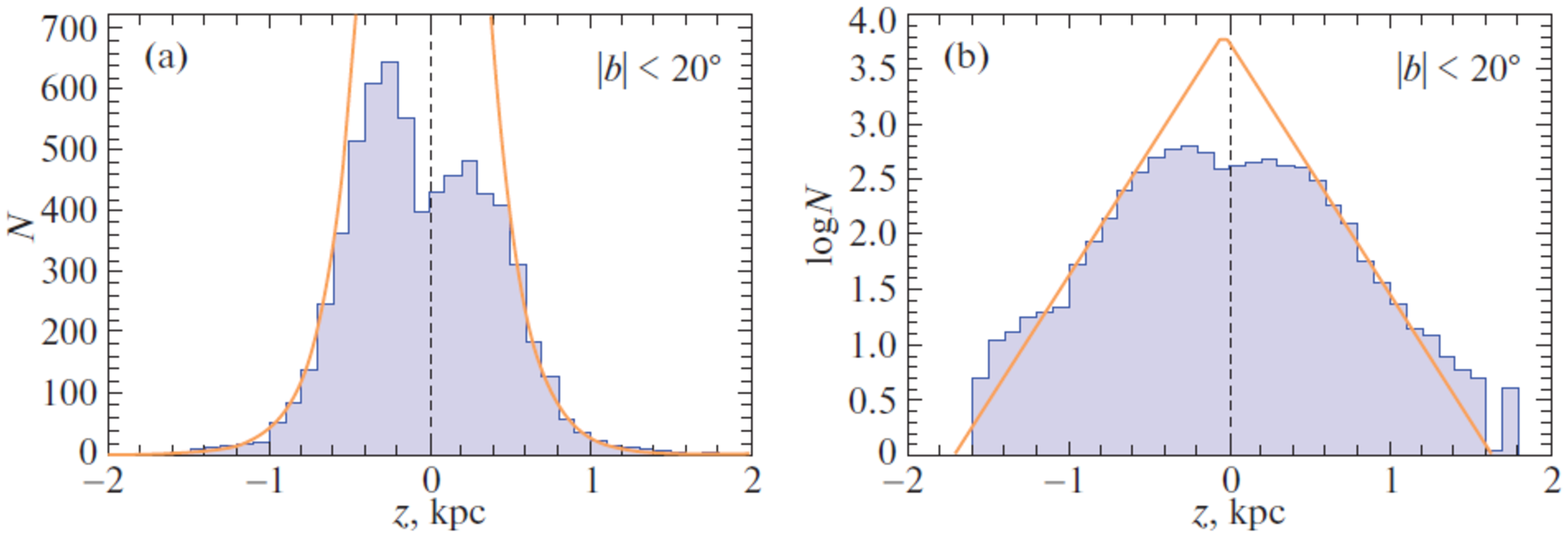}
 \caption{
Histogram of the $z$ distribution for the low-latitude HSDs in
linear (a) and logarithmic (b) scales. }
 \label{f1}
 \end{center} } \end{figure}
 \begin{figure} {\begin{center}
 \includegraphics[width=140mm]{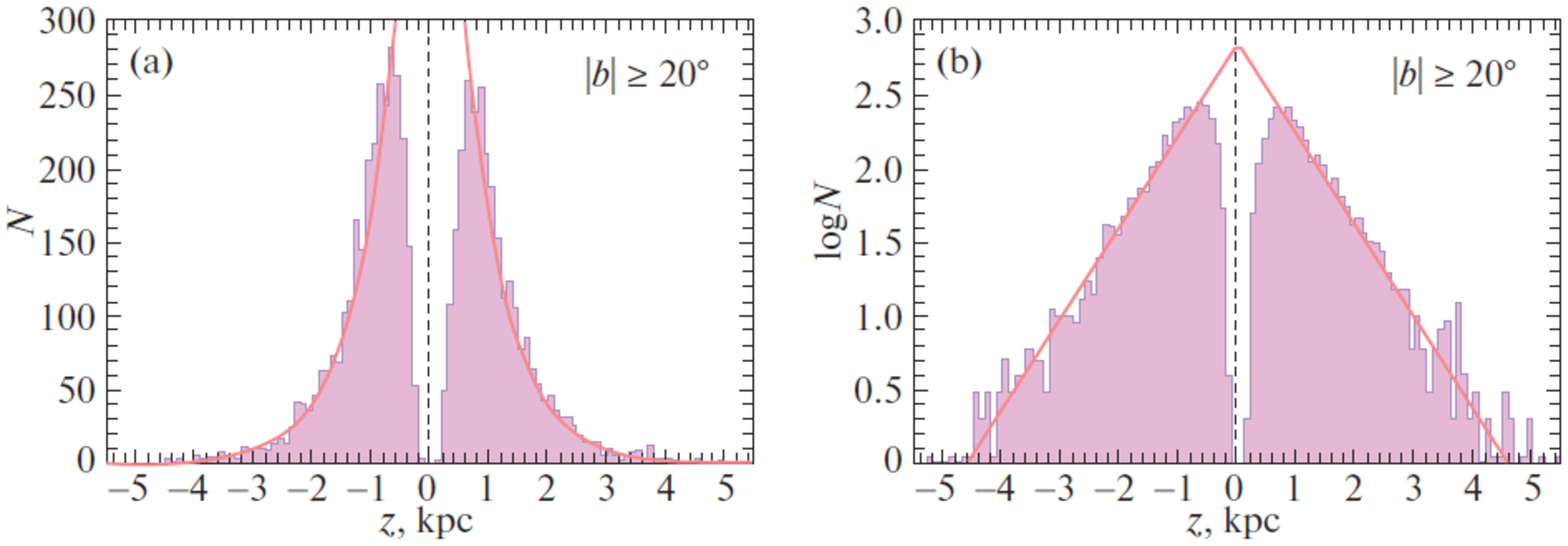}
 \caption{
Histogram of the $z$ distribution for the hight-latitude HSDs in
linear (a) and logarithmic (b) scales. }
 \label{f2}
 \end{center} } \end{figure}

Based on such stars, but with relative trigonometric parallax
errors less than 30\% (6094 stars), we found the following
residual velocity dispersions:
 \begin{equation}
 \begin{array}{lll}
  \sigma_1=56.7\pm0.9~\hbox{km s$^{-1}$}, \\
  \sigma_2=54.3\pm1.3~\hbox{km s$^{-1}$}, \\
  \sigma_3=39.8\pm0.6~\hbox{km s$^{-1}$}
 \label{rezult-6}
 \end{array}
 \end{equation}
and the orientation parameters of this ellipsoid
 \begin{equation}
  \matrix {
  L_1=~35\pm14^\circ, & B_1=-6\pm3^\circ, \cr
  L_2=125\pm39^\circ, & B_2=-5\pm2^\circ, \cr
  L_3=~74\pm39^\circ, & B_3=82\pm2^\circ. \cr
   }
 \label{rezult-66}
 \end{equation}
Apart from the clear difference in the sizes of the principal
semiaxes of the ellipsoids (19)--(21) and (23)--(25), we can note
a significant difference in the orientation of the third axis of
the ellipsoid for the high-latitude HSDs. This axis $(B_3)$ is
significantly deflected from the direction to the Galactic Pole by
$15\pm4^\circ$ (24); such an inclination is also confirmed by the
direction of the second axis ($B_2=-15\pm2^\circ$) of this
ellipsoid.

It is interesting to compare our estimates, for example, with the
results of the analysis of white dwarfs. Pauli et al. (2006)
derived the following velocity dispersions:
 $(\sigma_U,\sigma_V,\sigma_W)=(34,24,18)$ km s$^{-1}$ for a sample of
361 thin-disk white dwarfs,
 $(\sigma_U,\sigma_V,\sigma_W)=(79,36,46)$ km s$^{-1}$ for
a sample of 27 thick-disk white dwarfs, and
 $(\sigma_U,\sigma_V,\sigma_W)=(138,95,47)$ km s$^{-1}$ for a sample of
7 halo white dwarfs. Note that  $(\sigma_U,\sigma_V,\sigma_W)$ are
the residual velocity dispersions directed along the $(x,y,z)$
coordinate axes. The directions of the principal axes of the
ellipsoids for the low-latitude HSDs (20)--(22) virtually coincide
with those of the $(x,y,z)$ coordinate axes. The same can also be
said about the orientation of the ellipsoid (24). We can see good
agreement between the axes of the ellipsoids for the low-latitude
HSDs (19)--(21) and those for the thin-disk white dwarfs. At the
same time, the axes of the ellipsoids for the high-latitude HSDs
(23)--(25) are slightly smaller than those for the thick-disk
white dwarfs from Pauli et al. (2006).

Note the paper by Rowell and Hambly (2011), where the various
properties of the thin and thick disks as well as the halo are
discussed in sufficient detail. In particular, for the exponential
law (15) they think the scale height $h=250$ kpc to be typical for
the thin disk. In their Table 4 these authors also gave the
following parameters:
 $(U,V,W)_\odot$=(8.62,20.04,7.10) km s$^{-1}$ and
 $(\sigma_U,\sigma_V,\sigma_W)$= (32.4, 23.0, 18.1) km s$^{-1}$ for thin-disk stars,
 $(U,V,W)_\odot$=(11.0,42.0,12.0) km s$^{-1}$ and
 $(\sigma_U,\sigma_V,\sigma_W)$= (50.0, 56.0,34.0) km s$^{-1}$ for thick disk stars,
 $(U,V,W)_\odot$=(26, 199, 12) km s$^{-1}$ and
 $(\sigma_U,\sigma_V,\sigma_W)$= (141.0, 106.0, 94.0) km s$^{-1}$ for halo stars.
These estimates were obtained by Fuchs et al. (2009) for the thin
disk based on a sample of M stars from the SDSS catalogue and by
Chiba and Beers (2000) for the thick disk and the halo based on
metal-poor stars in the solar neighborhood. Our results are close
to the corresponding values for the thin and thick disks.

Bobylev and Bajkova (2017a) found the following principal semiaxes
of the residual velocity dispersion tensor:
$(\sigma_1,\sigma_2,\sigma_3)=(47,41,29)$ km s$^{-1}$ from a
sample of relatively young protoplanetary nebulae (with
luminosities higher than $5000 L_\odot$),
$(\sigma_1,\sigma_2,\sigma_3)=(50,38,28)$ km s$^{-1}$ from a
sample of older protoplanetary nebulae (with luminosities of $4000
L_\odot$ or $3500 L_\odot$), and, finally,
$(\sigma_1,\sigma_2,\sigma_3)=(91,49,36)$ km s$^{-1}$ from the
oldest nebulae belonging to the halo (with luminosities of $1700
L_\odot$).

Having analyzed the proper motions and parallaxes of stars from
the Gaia DR1 catalogue (Brown et al. 2016), Anguiano et al. (2018)
found the following dispersions:
 $(\sigma_U,\sigma_V,\sigma_W)=(33,28,23)\pm(4,2,2)$ km s$^{-1}$ and
 $(\sigma_U,\sigma_V,\sigma_W)=(57,38,37)\pm(6,5,4)$ km s$^{-1}$ for
the thin and thick disks, respectively. They showed that for
various thin-disk stellar groupings the vertex deviation
$(l_{uv})$ in the $UV$ plane changes in a very wide range, from
$-5^\circ$ to $+40^\circ,$ while the tilt angle $(l_{uw})$ in the
$UW$ plane varies from $-10^\circ$ to $+15^\circ$.

We may conclude that the low- and high-latitude HSDs exhibit the
thin- and thick-disk kinematics, respectively. Of course, the
presence of halo stars in the samples considered must not be ruled
out. However, their influence on the kinematics is statistically
negligible.

 \begin{figure} {\begin{center}
 \includegraphics[width=140mm]{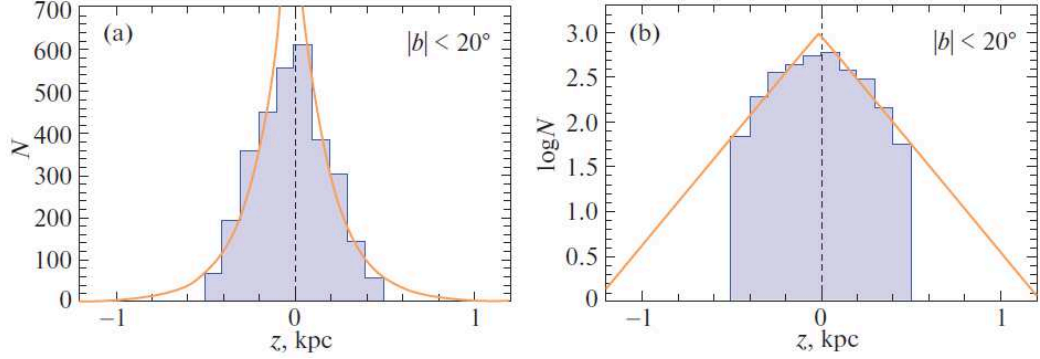}
 \caption{
Histogram of the $z$ distribution for the low-latitude HSDs
selected under the condition $r<1.5$ kpc in linear (a) and
logarithmic (b) scales. }
 \label{f3}
 \end{center} } \end{figure}
 \begin{figure} {\begin{center}
 \includegraphics[width=140mm]{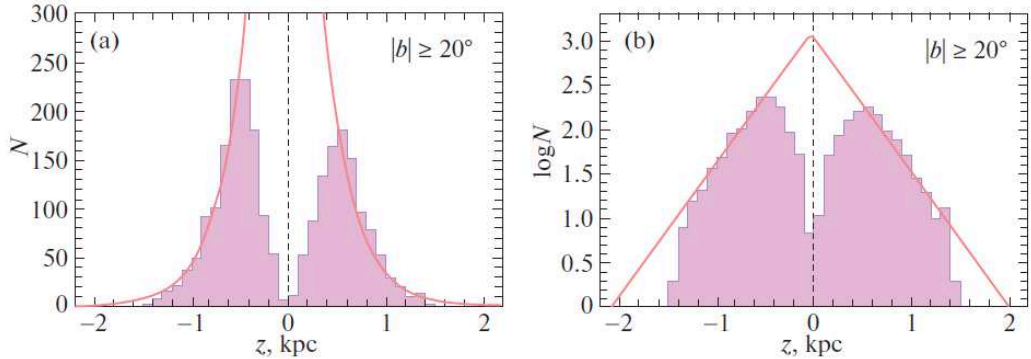}
 \caption{
Histogram of the $z$ distribution for the high-latitude HSDs
selected under the condition $r<1.5$ kpc in linear (a) and
logarithmic (b) scales.  }
 \label{f4}
 \end{center} } \end{figure}

 \subsection*{Parameters of the Density Distribution}
The values of $z_\odot$ and $h$ for the low- and high-latitude
HSDs used below to construct the exponential distribution (15) are
given in the upper part of Table 2. The results of the analysis of
various Galactic subsystems obtained by other authors are given in
the middle part of this table.

Sample 1 in Bobylev and Bajkova (2017a) consists of relatively
young protoplanetary nebulae with luminosities higher than
5000$L_\odot$, while sample 2 includes older protoplanetary
nebulae with luminosities of 4000$L_\odot$ or 3500$L_\odot$. As
can be seen from Table 2, the estimate of $h=190\pm4$ pc obtained
from the low-latitude HSDs is close to the value found from
planetary nebulae (Bobylev and Bajkova 2017b). In contrast, the
estimate of $h=700\pm8$ obtained from the high-latitude HSDs
exceeds the value found from sample 2 of protoplanetary nebulae
(Bobylev and Bajkova 2017a).

Figures 1 and 2 present histograms of the $z$ distribution for the
low- and high-latitude HSDs, respectively. Note the difference in
the sizes of the horizontal axes in these figures. A deficit of
stars near $z=0$ is seen in the figures. In Fig. 2 this dip is
largely due to the applied condition $|b|\geq20^\circ$, while the
conditions for the selection of candidate stars in the catalogue
by Geier et al. (2019) are responsible for the ``cut-off'' top of
the distribution in Fig. 1.

 \begin{table}[t]
 \caption[]{\small
Parameters of the residual velocity ellipsoid derived from a
complete sample
 }
  \begin{center}  \label{t:03}  \small
  \begin{tabular}{|l|r|r|r|r|r|}\hline
 Parameters             & \multicolumn{2}{|c|}{$|b|<20^\circ$} & \multicolumn{2}{|c|}{$|b|\geq20^\circ$} \\\hline
 $\sigma_\pi/\pi$       &        $<$15\% &         $<$30\% &        $<$15\% &         $<$30\% \\
 $N_\star$              &           2938 &            3761 &           2176 &           2920 \\
 ${\overline r},$ kpc   &           0.96 &            0.97 &           0.98 &           0.96 \\
 $\overline {|z|},$ kpc &           0.16 &            0.17 &           0.57 &           0.56 \\
 $\sigma_1,$ km s$^{-1}$ & $35.60\pm0.72$ & $34.91\pm0.58$  & $45.23\pm0.86$ & $43.78\pm0.74$ \\
 $\sigma_2,$ km s$^{-1}$ & $28.19\pm2.11$ & $26.83\pm1.74$  & $37.68\pm1.80$ & $37.70\pm1.85$ \\
 $\sigma_3,$ km s$^{-1}$ & $21.04\pm1.06$ & $20.76\pm0.74$  & $30.18\pm0.84$ & $27.89\pm0.69$ \\
       $L_1, B_1$     & $~10\pm12^\circ,$ $~3\pm3^\circ$ & $~~9\pm9^\circ,$ ~$~2\pm3^\circ$ & $~4\pm8^\circ,$  $~-1\pm2^\circ$ & $~12\pm9^\circ,$  $~-2\pm2^\circ$ \\
       $L_2, B_2$     & $100\pm4^\circ,$ ~$~8\pm3^\circ$ & $~99\pm3^\circ,$ ~$~6\pm2^\circ$ & $94\pm7^\circ,$  $~-1\pm2^\circ$ & $102\pm9^\circ,$  $~-6\pm2^\circ$ \\
       $L_3, B_3$     & $262\pm4^\circ,$ ~$81\pm5^\circ$ & $258\pm3^\circ,$ ~$83\pm4^\circ$ & $53\pm7^\circ,$ ~$~88\pm7^\circ$ & $~81\pm9^\circ,$ ~$~84\pm4^\circ$ \\
  \hline
 \end{tabular}\end{center}
  {\small $N_\star$ is the number of stars used, ${\overline r}$ is the mean
distance of the sample of stars.}
  \end{table}
 \begin{table}[t]
 \caption[]{\small
Parameters of the residual velocity ellipsoid found from a
complete sample
 }
  \begin{center}  \label{t:04}  \small
  \begin{tabular}{|l|r|r|r|r|r|}\hline
 Parameters             &  $|z|<0.2$~kpc & $|z|:0.2-0.4$~kpc & $|z|:0.4-0.6$~kpc & $|z|\geq0.6$~kpc \\\hline
 $N_\star$              &           2499 &            1971 &           1193 &           1114 \\
 ${\overline r},$ kpc   &           0.82 &            0.99 &           1.04 &           1.18 \\
 $\overline {|z|},$ kpc &         0.10 &            0.29 &           0.49 &           0.81 \\
 $U_\odot,$ km s$^{-1}$ & $ 8.2\pm0.7$ & $ 9.2\pm0.9$  & $ 9.6\pm1.3$ & $11.3\pm1.5$ \\
 $V_\odot,$ km s$^{-1}$ & $17.2\pm1.0$ & $23.4\pm1.4$  & $27.0\pm1.9$ & $41.3\pm2.1$ \\
 $W_\odot,$ km s$^{-1}$ & $ 6.1\pm0.5$ & $ 7.5\pm0.8$  & $ 6.7\pm1.2$ & $ 4.8\pm1.9$ \\

 $\Omega_0,$ km s$^{-1}$ kpc$^{-1}$     & $-30.0\pm0.9$ & $-27.7\pm1.1$ & $-26.7\pm1.7$ & $-23.9\pm2.3$ \\
 $\Omega^{'}_0,$ km s$^{-1}$ kpc$^{-2}$ & $  3.3\pm0.2$ & $  3.6\pm0.2$ & $  3.1\pm0.3$ & $  2.5\pm0.5$ \\
 $V_0,$ km s$^{-1}$                     &   $240\pm9$   &   $221\pm10$  &   $214\pm14$  &   $191\pm19$  \\

 $\sigma_1,$ km s$^{-1}$ & $34.0\pm0.8$ & $36.7\pm0.8$  & $41.7\pm1.5$ & $45.5\pm1.1$ \\
 $\sigma_2,$ km s$^{-1}$ & $24.2\pm2.9$ & $27.3\pm1.2$  & $33.6\pm4.8$ & $37.1\pm2.6$ \\
 $\sigma_3,$ km s$^{-1}$ & $19.2\pm1.3$ & $23.1\pm0.7$  & $28.2\pm1.4$ & $31.2\pm0.9$ \\

  $L_1, B_1$ & $~10\pm20^\circ,$  $~4\pm9^\circ$ & $~2\pm2^\circ,$ $~0\pm1^\circ$ & $ ~1\pm20^\circ,$ $~1\pm5^\circ$~ & $8\pm20^\circ,$ $-2\pm6^\circ$ \\
  $L_2, B_2$ & $100\pm3^\circ,$   $~9\pm3^\circ$ & $92\pm4^\circ,$ $~8\pm2^\circ$ & $~91\pm16^\circ,$ $-2\pm5^\circ$~ & $98\pm7^\circ,$ $-1\pm3^\circ$ \\
  $L_3, B_3$ & $257\pm3^\circ,$   $80\pm7^\circ$ & $91\pm4^\circ,$ $82\pm8^\circ$ & $116\pm16^\circ,$ $87\pm17^\circ$ & $40\pm7^\circ,$ $88\pm9^\circ$ \\
  \hline
 \end{tabular}\end{center}
 {\small $N_\star$ is the number of stars used, ${\overline r}$ is the mean
distance of the sample of stars. }
\end{table}

 \subsection*{Estimating the Surface Density}
A complete stellar sample is required to estimate the surface
density. According to the estimates by Geier et al. (2019), except
for the narrow region near the Galactic plane, the catalogue of
HSDs is complete to a distance of 1.5 kpc, which we take as the
completeness boundary. Thus, in this section we consider the low-
and high-latitude HSDs selected under the condition $r<1.5$ kpc.
The results of their analysis are presented in Tables 2 and 3 as
well as in Figs. 3 and 4.

As can be seen from Table 2, the scale height $h$ remains almost
constant for the low-latitude HSDs under various sample production
conditions. In contrast, $h$ for the high-latitude HSDs h depends
strongly on the constraints used. We can see that the moderate
$h=290\pm10$ kpc is consistent with the results of the analysis of
white dwarfs (Vennes et al. 2002) and planetary nebulae (Bobylev
and Bajkova 2017b).

As can be seen from Table 3, the third axis in all ellipsoids is
close to the direction to the Galactic Pole. Therefore, in Eq.
(17) we may set $\overline {v^2_z}=\sigma^2_3$. This is apparently
the only way of estimating the vertical velocity, because in our
case we have no information about the radial velocities and,
therefore, we cannot directly calculate the space velocities $U,
V,$ and $W.$

Let us make an estimate for the mean $|z|$ given in Table 3. We
see that $\overline {|z|}=0.17$ kpc for the low-latitude HSDs,
which is of no real interest for the estimate of $\Sigma_{out}$
due to the small $z.$ From the high-latitude HSDs at
$\sigma_\pi/\pi<30\%$ and $z_{out}=\overline {|z|}=0.56$ kpc we
find $\Sigma_{out}=53\pm4~M_{\odot}$ kpc$^{-2}$.

For comparison, we can give some values of the surface density
obtained by other authors. Korchagin et al. (2003) found
$\Sigma_{out}=46\pm2~M_{\odot}$ kpc$^{-2}$ for $z_{out}=0.35$ kpc
based on a sample of $\sim$1500 red giants from the Hipparcos
catalogue (1997). Based on giants from the Hipparcos catalogue
(1997), Holmberg and Flynn (2004) estimated the local density of
the visible matter, $\Sigma_{out}=56\pm6~M_{\odot}$ kpc$^{-2}$,
and the entire gravitating matter, including the dark matter disk
and halo, $\Sigma_{out}=74\pm6~M_{\odot}$ kpc$^{-2}$, for
$z_{out}=$1.1 kpc. Based on $\sim$16 000 G dwarfs from the SEGUE
catalogue (Yanny et al. 2007), Bovy and Rix (2013) estimated the
local stellar density, $\Sigma_{out}=38\pm4~M_{\odot}$ kpc$^{-2}$,
and the density of the entire gravitating matter,
$\Sigma_{out}=68\pm4~M_{\odot}$ kpc$^{-2}$, for $z_{out}=$1.1 kpc.
Based on stars from the TGAS catalogue (Tycho--Gaia Astrometric
Solution, Prusti et al. 2016; Lindegren et al. 2016), Kipper et
al. (2018) found $\Sigma_{out}=42\pm4~M_{\odot}$ kpc$^{-2}$ for
$z_{out}\leq0.75$ kpc.

 \subsection*{Dependence of Kinematic Parameters on $|z|$}
In this section we consider the dependence of some kinematic
parameters on the coordinate $|z|$ based on a complete sample of
HSDs. For this purpose, we used a sample of HSDs with errors
$\sigma_\pi/\pi$<30\%, which we divided into four nonoverlapping
parts, depending on $|z|.$ The results are presented in Table 4.
Equations (1)--(2) were solved with five unknowns: $U_\odot,
V_\odot, W_\odot, \Omega_0,$ and $\Omega^{'}_0$. Here we did not
divide the stars into low- and high-latitude ones. We see good
agreement between the parameters of the residual velocity
ellipsoid inferred from both high-latitude HSDs (Table 3) and HSDs
with large $|z|$ (Table 4). In particular, $\sigma_3\rightarrow30$
km s$^{-1}$ at large $|z|.$ Thus, we correctly estimated the
surface density $\Sigma_{out}.$

Based on the data from this table, we estimated the gradient of
the circular rotation velocity $V_0$ as a function of $|z|,$
${\displaystyle\Delta V_0\over\displaystyle \Delta |z|}=-64\pm5$
km s$^{-1}$ kpc$^{-1}$, by fitting a linear trend from four
measurements. For comparison, based on a sample of metal-poor
stars in the solar neighborhood, Chiba and Beers (2000) obtained
the following estimates of such gradients: ${\displaystyle\Delta
V_0\over\displaystyle \Delta |z|}=-30\pm3$ km s$^{-1}$ kpc$^{-1}$
for thin-disk stars with a velocity ellipsoid
$(\sigma_U,\sigma_V,\sigma_W)=(46,50,35)\pm(4,4,3)$ km s$^{-1}$
and ${\displaystyle\Delta V_0\over\displaystyle \Delta
|z|}=-52\pm6$ km s$^{-1}$ kpc$^{-1}$ for halo stars with a highly
elongated velocity ellipsoid
$(\sigma_U,\sigma_V,\sigma_W)=(141,106,94)\pm(11,9,8)$ km
s$^{-1}$.

 \section*{CONCLUSIONS}
We studied the kinematics of hot dim stars from the catalogue by
Geier et al. (2019). These are 39 800 HSD candidates selected by
them from the Gaia DR2 catalogue using data from several multiband
photometric sky surveys. In this paper we analyzed more than 12
500 proper motions of stars with relative trigonometric parallax
errors less than 30\%. Thus,we investigated a huge set of
present-day highly accurate data that is of great value for a
statistical analysis.

The stars were divided into two parts, depending on the Galactic
latitude, into low-latitude ($|b|<20^\circ$) and high-latitude
($|b|\geq20^\circ$) samples. Such a breakdown is needed to
construct the histograms in a wide $z$ range and to determine the
parameters of the exponential density distribution from them.
These two samples were shown to have completely different
kinematics. We determined the Galactic rotation parameters from
relatively more distant objects whose parallax errors do not
exceed 30\%. In that case, at least the first derivative of the
angular velocity of Galactic rotation is determined quite
reliably. In contrast, for the subsequent determination of the
residual velocity ellipsoid parameters we already use more
reliable data, objects with trigonometric parallax errors no
greater than 15\%.

For example, the sample of low-latitude HSDs rotates around the
Galactic center with a linear velocity $V_0=221\pm5$ km s$^{-1}$.
This suggests that they belong to the Galactic thin disk. They lag
behind the local standard of rest by $\Delta V_\odot\sim16$ km
s$^{-1}$ due to the asymmetric drift. The high-latitude HSDs
rotate with a considerably lower velocity, $V_0=168\pm6$ km
s$^{-1}$, which is more typical for the thick disk. Their lagging
behind the local standard of rest is $\Delta V_\odot\sim40$ km
s$^{-1}$.

A joint analysis of the entire sample of 12 515 stars with
parallax errors less than 30\% revealed a positive rotation around
the x axis differing significantly from zero with an angular
velocity $\omega_1=1.36\pm0.24$ km s$^{-1}$ kpc$^{-1}$. Expressed
in different units, given the mean distance of the sample stars,
this angular velocity is $\omega_1=0.15\pm0.03$3 mas yr$^{-1}$.
The latter value is consistent with the estimates by Lindblad et
al. (2018) for the residual rotation of the Gaia DR2 reference
frame relative to the inertial reference frame.

Based on a sample of 4181 low-latitude HSDs with parallax errors
less than 15\%, we determined the following principal semiaxes of
the residual velocity ellipsoid:
$(\sigma_1,\sigma_2,\sigma_3)=(37.4,28.1,22.8)\pm(0.9,0.7,0.9)$ km
s$^{-1}$. To within 3--4$^\circ$, the principal axes of this
ellipsoid coincide in directions with the axes of the Galactic
rectangular coordinate system $x, y, z.$

Based on a sample of 3584 high-latitude HSDs with parallax errors
less than 15\%, we found
$(\sigma_1,\sigma_2,\sigma_3)=(51.9,46.6,34.8)\pm(1.1,1.8,0.8)$ km
s$^{-1}$, with the third axis of this ellipsoid (24) being
deflected from the direction to the Galactic Pole by
$15\pm4^\circ.$ Such a deflection is also confirmed by the
direction of the second axis $B_2=-15\pm2^\circ,$ with the error
having been estimated here along each axis independently.

We studied two samples of HSDs that are complete within $r<1.5$
kpc. The following vertical disk scale heights were derived:
$h=180\pm6$ pc and $290\pm10$ pc from the low- and high-latitude
HSDs, respectively. Here, we obtained a new estimate of the local
stellar density $\Sigma_{out}=53\pm4~M_{\odot}$ kpc$^{-2}$ for
$z_{out}=0.56$ kpc from the high-latitude HSDs.

Based on a sample that is complete with errors
$\sigma_\pi/\pi$<30\%, we traced the evolution of the kinematic
parameters as a function of the coordinate $|z|.$ We estimated the
gradient of the circular rotation velocity $V_0$ along $|z|,$
${\displaystyle\Delta V_0\over\displaystyle\Delta|z|}=-64\pm5$ km
s$^{-1}$ kpc$^{-1}$.

All of the above results lead us to conclude that the low- and
high-latitude HSDs exhibit the kinematics of the thin and thick
disks, respectively. This suggests that most of the low-latitude
HSDs ``inherited'' the kinematic properties of relatively young
massive precursor stars. The high-latitude HSDs are less
homogeneous kinematically; their properties depend strongly on the
positions above the Galactic plane and the heliocentric distance.
Kinematically more ``heated'' objects that have already receded
quite far from the Galactic plane are apparently their precursors.

 \section*{ACKNOWLEDGMENTS}
We are grateful to the referees for the useful remarks that
contributed to an improvement of the paper.

 \bigskip \bigskip\medskip{\bf REFERENCES}{\small

 1. M. Altmann, H. Edelmann, and K. S. de Boer, Astron. Astrophys. 414, 181 (2004).

 2. B. Anguiano, S. R. Majewski, K. C. Freeman, A. W. Mitschang, and
M. C. Smith, Mon. Not. R. Astron. Soc. 474, 854 (2018).

 3. F. Arenou, X. Luri, C. Babusiaux, C. Fabricius, A. Helmi, T.
Muraveva, A. C. Robin, F. Spoto, et al. (Gaia Collab.), Astron.
Astrophys. 616, 17 (2018).

  4. V. V. Bobylev, Astron. Lett. 36, 634 (2010).

 5. V. V. Bobylev, Astron. Lett. 39, 819 (2013).

 6. V. V. Bobylev and A. T. Bajkova, Astron. Lett. 40, 389 (2014).

 7. V. V. Bobylev and A. T. Bajkova, Astron. Lett. 42, 1 (2016).

 8. V. V. Bobylev and A. T. Bajkova, Astron. Lett. 43, 452 (2017a).

 9. V. V. Bobylev and A. T. Bajkova, Astron. Lett. 43, 304 (2017b).

 10. V. V. Bobylev and A. T. Bajkova, Astron. Lett. 45, 208 (2019).

 11. C. Bonatto, L. O. Kerber, E. Bica, and B. X. Santiago, Astron.
Astrophys. 446, 121 (2006).

 12. J. Bovy and H.-W. Rix, Astrophys. J. 779, 115 (2013).

 13. A. G. A. Brown, A. Vallenari, T. Prusti, J. de Bruijne, F. Mignard,
  R. Drimmel, et al. (Gaia Collab.), Astron. Astrophys. 595, 2 (2016).

 14. A. G. A. Brown, A. Vallenari, T. Prusti, de Bruijne, C.
Babusiaux, C. A. L. Bailer-Jones, M. Biermann, D.W. Evans, et al.
(Gaia Collab.), Astron. Astrophys. 616, 1 (2018).

 15. Y. Bu, Z. Lei, G. Zhao, J. Bu, and J. Pan, Astrophys. J.
Suppl. Ser. 233, 2 (2017).

 16. T. Camarillo, M. Varun, M. Tyler, and R. Bharat, Publ. Astron.
Soc. Pacif. 130, 4101 (2018).

 17. M. Chiba and T. C. Beers, Astron. J. 119, 2843 (2000).

 18. G. Fontaine, P. Brassard, S. Charpinet, E. M. Green, S. K.
Randall, and V. Van Grootel, Astron. Astrophys. 539, 12 (2012).

 19. B. Fuchs, C. Dettbarn, H.-W. Rix, T. C. Beers, D. Bizyaev, H.
Brewington, H. Jahreiss, R. Klement, et al., Astron. J. 137, 4149
(2009).

 20. S. Geier, S. Nesslinger, U. Heber, N. Przybilla, R.
Napiwotzki, and R.-P. Kudritzki, Astron. Astrophys. 464, 299
(2007).

 21. S. Geier, T. Kupfer, U. Heber, V. Schaffenroth, B. N. Barlow,
R. H. Oestensen, S. J. O'Toole, E. Ziegerer, et al., Astron.
Astrophys. 577, 26 (2015).

 22. S. Geier, R. Raddi, N. P. Gentile Fusillo, and T. R. Marsh,
Astron. Astrophys. 621, 38 (2019).

 23. J. L. Greenstein and A. I. Sargent, Astrophys. J. Suppl. 28, 157 (1974).

 24. R. de Grijs and G. Bono, Astrophys. J. Suppl. Ser. 232, 22 (2017).

 25. Z. Han, Ph. Podsiadlowski, and A. E. Lynas-Gray, Mon. Not. R.
Astron. Soc. 380, 1098 (2007).

 26. The HIPPARCOS and Tycho Catalogues, ESA SP--1200 (1997).

 27. J. Holmberg and C. Flynn, Mon. Not. R. Astron. Soc. 352, 440
(2004).

 28. M. L. Humason and F. Zwicky, Astrophys. J. 105, 85 (1947).

 29. R. Kipper, E. Tempel, and P. Tenjes, Mon. Not. R. Astron. Soc.
473, 2188 (2018).

 30. V. I. Korchagin, T. M. Girard, T. V. Borkova, D. I. Dinescu,
and W. F. van Altena, Astron. J. 126, 2896 (2003).

 31. T. Kupfer, S. Geier, U. Heber, R. H. Ostensen, B. N. Barlow,
P. F. L. Maxted, C. Heuser, V. Schaffenroth, and B. T. Gansicke,
Astron. Astrophys. 576, 44 (2015).

  32. M. Latour, P. Chayer, E. M. Green, A. Irrgang, and G.
Fontaine, Astron. Astrophys. 609, 89 (2018).

 33. Z. Lei, G. Zhao, A. Zeng, L. Shen, Z. Lan, D. Jiang, and Z.
Han, Mon. Not. R. Astron. Soc. 463, 3449 (2016).

 34. L. Lindegren, U. Lammers, U. Bastian, J. Hernandez, S. Klioner,
D. Hobbs, A. Bombrun, D. Michalik, et al., Astron. Astrophys. 595,
4 (2016).

 35. L. Lindegren, J. Hernandez, A. Bombrun, S. Klioner, U.
Bastian, M. Ramos-Lerate, A. de Torres, H. Steidelmuller, et al.
(Gaia Collab.), Astron. Astrophys. 616, 2 (2018).

 36. M. L\'opez-Corredoira, H. Abedi, F. Garz\'on, and F. Figueras,
Astron. Astrophys. 572, 101 (2014).

37. M. Miyamoto and Z. Zhu, Astron. J. 115, 1483 (1998).

 38. K.F. Ogorodnikov, {\it Dynamics of stellar systems} (Oxford: Pergamon, ed. Beer, A. 1965).

 39. E.-M. Pauli, R. Napiwotzki, U. Heber, M. Altmann, and M.
Odenkirchen, Astron. Astrophys. 447, 173 (2006).

 40. T. Prusti, J.H. J. de Bruijne, A. G. A. Brown, A. Vallenari, C.
Babusiaux, C. A. L. Bailer-Jones, U. Bastian, M. Biermann, et al.
(Gaia Collab.), Astron. Astrophys. 595, 1 (2016).

 41. S. K. Randall, A. Calamida, G. Fontaine, M. Monelli, G.
Bono, M. L. Alonso, V. Van Groote, P. Brassard, et al., Astron.
Astrophys. 589, 1 (2016).

 42. A. S. Rastorguev, M. V. Zabolotskikh, A. K. Dambis, N. D.
Utkin, V. V. Bobylev, and A. T. Bajkova, Astrophys. Bull. 72, 122
(2017).

 43. A. G. Riess, S. Casertano, W. Yuan, L. Macri, B. Bucciarelli, M.
G. Lattanzi, J. W. MacKenty, J. B. Bowers, et al., Astrophys. J.
861, 126 (2018).

 44. M. Romero-G\'omez, C. Mateu, L. Aguilar, F. Figueras, and A.
Castro-Ginard, arXiv: 1812.07576 (2018).

 45. N. Rowell and N. C. Hambly, Mon. Not. R. Astron. Soc. 417, 93 (2011).

 46. R. Sch\"onrich, J. Binney, and W. Dehnen, Mon. Not. R. Astron. Soc.
403, 1829 (2010).

 47. L. Spitzer, Astrophys. J. 95, 329 (1942).

 48. K. G. Stassun and G. Torres, Astrophys. J. 862, 61 (2018).

  49. A. S. Tsvetkov and F. A. Amosov, Astron. Lett. 45, 462 (2019).

 50. J. P. Vall\'ee, Astrophys. Space Sci. 362, 79 (2017).

 51. S. Vennes, R. J. Smith, B. J. Boyle, S. M. Croom, A. Kawka, T.
Shanks, L. Miller, and N. Loaring, Mon. Not. R. Astron. Soc. 335,
673 (2002).

 52. V. V. Vityazev, A. S. Tsvetkov, V. V. Bobylev, and A. T.
Bajkova, Astrofizika 60, 503 (2017).

 53. J. Vos, P. N\'emeth, M. Vu\u {c}kovi\'c, R. Ostensen, and S.
Parsons, Mon. Not. R. Astron. Soc. 473, 693 (2018).

 54. L. N. Yalyalieva, A. A. Chemel’, E. V. Glushkova, A. K.
Dambis, and A. D. Klinichev, Astrophys. Bull. 73, 335 (2018).

55. B. Yanny, C. Rockosi, H. J. Newberg, G. R. Knapp, J. K.
Adelman-McCarthy, B. Alcorn, S. Allam, C. A. Prieto, et al.,
Astron. J. 137, 4377 (2009).

56. J. C. Zinn, M. H. Pinsonneault, D. Huber, and D. Stello,
arXiv: 1805.02650 (2018).
  }
  \end{document}